\documentclass[preprint,12pt]{aastex}

\begin{document}

\title {High-Resolution {\it Chandra} 
Spectroscopy of $\gamma$ Cassiopeia (B0.5e)}

\author{Myron A. Smith} 
\affil {Catholic University of America \& Computer Sciences Corporation/STScI, 
3700 San Martin Dr., Baltimore, MD 21218; msmith@stsci.edu}
\author{David H. Cohen} 
\affil{Department of Physics \& Astronomy, Swarthmore College, Swarthmore, 
PA 19081} 
\author{Ming Feng Gu\footnote{Chandra Fellow} }
\affil{Center for Space Research, Massachusetts Institute of Technology, 
Cambridge, MA, 02139}
\author{ Richard D. Robinson} 
\affil {Catholic University of America \& Computer Sciences Corporation/Johns
Hopkins University } 
\author{Nancy Remage Evans} 
\affil{ Center for Astrophysics, Harvard University, Cambridge, MA 02138} 
\author{Prudence G. Schran }
\affil{Department of Physics \& Astronomy, Swarthmore College, Swarthmore, 
PA 19081} 

\newpage

\begin{abstract}

  $\gamma$\,Cas is the prototypical classical B0.5e star and is now 
known to be the primary in a wide binary system. It has long been famous 
for its unique hard X-ray characteristics, among which are variations that 
correlate with changes in a number of optical light and UV-line and 
continuum properties. These peculiarities have led to a picture in which 
processes on or near the Be star produce the observed X-ray emission. 
In this paper we report on a 53\,ks {\it Chandra} HETGS observation of 
this target. 

   An inspection of our spectrum shows that it is quite atypical 
for a massive star. The emission lines appear ``weak" because of a strong 
short-wavelength continuum that arises from a hot plasma with
k{\it T} = 11\,--\,12 keV. The spectrum exhibits many lines, the strongest of 
which are Ly$\alpha$ 
features of H-like species from Fe through the even-Z, intermediate elements 
(S, Si, Mg, Ne) down to O and N. Line ratios of the ``$rif$ triplet" 
for a variety of He-like ions 
and of \ion{Fe}{17}  are consistent with the dominance of collisional 
atomic processes. However, the presence of Fe and Si fluorescence K features 
indicates that photoionization also occurs in nearby cold gas.
The line profiles indicate a mean velocity at rest with a r.m.s. line 
broadening of 500 km\,s$^{-1}$ and little or no asymmetry.  An empirical
global fitting analysis of the line and continuum spectrum suggests that
there are actually 3--4 plasma emission components. 
The first is the dominant hot (12\,keV) component, of which some fraction 
(10\,--\,30\%) is heavily absorbed, while the remainder is affected by
only  a much lower column density of 3$\times10^{21}$ cm$^{-2}$. 
The hot component has a Fe abundance of only 0.22$\pm{.05}$ solar. 
The other two or three major emission components are 
``warm" and are responsible for most other emission lines. These
components are dominated by plasma having temperatures near 0.1, 0.4, and 
3\,keV. Altogether, the warm components have
an emission measure of about 14\% of the hot component, a low column 
density, and a more nearly solar composition. The 100 eV component
is consistent with X-ray temperatures associated with a wind in a
typical early B star. Nonetheless, its emission measure is a few times higher 
than would be expected by this explanation.
The strength of the fluorescence features and the dual-column 
absorption model for the hot plasma component suggest the presence near 
the hot sites of a cold gas structure with a column density of $\sim10^{23}$ 
cm$^{-2}$. Because this is also the value determined by Millar \& Marlborough 
for the vertical column of the Be disk of $\gamma$\,Cas, these attributes 
suggest that the X-ray emitting sources could be close to the disk and
hence the Be star. 
Finally, we discuss the probably related issues of the origin of 
the warm emission components as well as the puzzling deficient Fe abundance 
in the hot component. It is possible that the latter anomaly is related 
to the FIP (abundance fractionation) effect found in 
certain coronal structures on the Sun and RS\,CVn stars. This would be
yet another indication that the X-rays are produced in the immediate
vicinity of the Be star.

\end{abstract}


\newpage

\section{Introduction }
\label{intro}

  $\gamma$\,Cas (B0.5\,IV) has held a unique place among the broad group 
of X-ray emitting OB stars because of its extensive and unusually dense
($n_{\rm e}$ $\simeq$ 10$^{13}$\,cm$^{-3}$; Waters et al. 1987) decretion
disk and the attributes of its X-ray emission. The disk has
been imaged in H$\alpha$ out to several stellar radii, and its
ellipsoidal shape permits an estimate of the orientation angle of the disk 
and rotation plane with respect to the line of sight, $\simeq$46$^{o}$ 
(Quirrenbach et al. 1996). Its various X-ray properties are peculiar and 
virtually unique for a massive star. First, its
X-ray luminosity (0.4--1.1$\times10^{33}$ ergs~s$^{-1}$) is midway
between the {\it L}$_{x}$ values of normal B and classical Be stars on the low
side and of X-ray Be binaries on the high side. The X-rays are also thermal,
but with an extremely high temperature of k{\it T} = 10.5--12\,keV. Several
investigators have attempted to explain this property by wind or
Be-disk infall onto a degenerate companion, generally a white dwarf (see Kubo
et al. 1998, Owens et al. 1999, Apparao 2002). Potentially, this picture can
explain in a general way the hard thermal spectrum of the X-rays. However,
while the star is now established as part of a binary system  (see
Harmanec et al. 2000; Miroshnichenko,  Bjorkman, \& Krugov 2002), the wide
separation ({\it P} $\approx$ 204 days, with low to moderate eccentricity) 
makes it difficult to understand the high $L_{\rm x}$ if the companion 
were a white dwarf. Moreover,
for a companion to have evolved to a degenerate star, it must have initially
been more massive than $\gamma$ Cas and would more likely have developed
into a neutron star than a white dwarf. A neutron star system can easily
explain the $L_{\rm x}$, but not the presence of dominant thermal processes
implied by the continuum shape
and the presence of \ion{Fe}{25} and \small{XXVI} lines. The spectra of Be-n.s.
binary systems generally tend to be nonthermal and show a strong fluorescence
feature near 6.5 keV, which is not seen in $\gamma$ Cas (Kubo et al, 1998).
In addition, neutron star systems generally have a strong tendency to have
eccentric orbits and to show periodic X-ray pulses, unlike $\gamma$\,Cas.

Smith, Robinson, \& Corbet (1998; SRC98) have shown that there are actually 
two components to the hard X-rays, each having about the same temperature.
One is characterized by rapid fluctuations (flares or ``shots") lasting from 
a few seconds (or less) to a few minutes. These must be emitted from
a high density, optically thin plasma. The second, ``basal" component
(also optically thin) varies on a timescale of hours and contributes
60\,--\,70\% of the total flux. SRC98 found that the basal X-ray variations
were anticorrelated with UV continuum variations near 1400 \AA\ and
proposed that the X-rays were actually coming from near $\gamma$ Cas itself.
In their model they proposed that the
shots were emitted by violent flares at the top of the photosphere
of the Be star. These exploding parcels expand with rather little energy
loss and fill a lower density cavity inside supposed magnetic loops
emerging from the star's surface, which is the site of the basal emission.
These filled cavities are thought to be associated with co-rotating clouds
of cool material which are responsible for striated subfeatures in 
time-series spectra of optical and UV line
profiles as well as  variations in the thermal properties of UV spectral lines
(Smith \& Robinson 1999, Cranmer, Smith, \& Robinson 2000; CSR00), and 
absorptions
in the UV continuum light curves (Smith, Robinson, \& Hatzes 1998; SRH98).
Highly redshifted spectral lines in the UV (Smith \& Robinson 1999) are
indicative of high velocity plasmoids, with energies comparable to the X-ray
flux and suggest the interaction of X-ray emitting volumes with a circumstellar
structure, probably the dense disk of the Be star.
There is indirect evidence from both UV/X-ray correlations (CSR00) and broad 
high-level hydrogen lines (Waters et al. 2000) that this interaction occurs 
in a region where circumstellar gas ceases to co-rotate at the angular rate
of the star's surface and begins to follow a Keplerian orbital relation.

  Recently, evidence for the star's disk being somehow associated with 
the X-ray generation entered the picture with the discovery by Robinson, 
Smith, \& Henry (2002; RSH02) of correlated optical/X-ray cycles with
a mean length of 70 days. The amplitudes of the variations are a factor
of three for the X-ray and 3\% for optical flux. Yet because of the small
contribution of the X-ray emission to the star's total luminosity, the 
optical variations cannot be produced by reprocessing of X-ray flux, and 
thus the optical optical variations must have another cause. The 
association of these variations with the disk is implied by their slightly 
reddish color, which probably means that they come from a source cooler 
than the Be star. Finally, the {\it cyclical} nature of the variations 
led RSH02 to suggest a dynamo origin. Thus, this work has potentially placed 
the generation of the X-rays into a comprehensive (though otherwise
untested) picture in which the X-ray and optical variations result from 
magnetic stresses between the star and the disk. 

   The X-ray properties of this star, and continued questions about the 
nature of its emission, have made it a natural target for spectroscopy. 
A high resolution spectrum permits a description of the 
temperatures, densities, and kinematics of the X-ray emitting 
regions. At a more 
qualitative level, an X-ray spectrum makes possible a comparison with 
the spectra of other well known systems, such as accreting
white dwarfs and neutron stars, and the determination of whether its 
spectral characteristics are as peculiar as its temporal and 
flux properties. To address these issues, we requested and were 
granted time with the High and Medium Energy Transmission
Gratings (HEG, MEG) of the {\it Chandra} satellite. We report herein 
on our analysis of this spectrum.

\section{ Observations and Reduction }

  Our HEG/MEG spectra were obtained with a 52.5 ks exposure time 
starting 2001 August 10, 9:21 UT.
We have reextracted the MEG and HEG spectra and created the Ancillary 
Response File and Response Matrix File using {\it CIAO} v2.3 and {\it
CALB} v2.18. This brought any shifts between the plus and minus spectral
orders below levels of detection
($\pm{0.2}$ pix), which was not the case in the original pipeline extraction.

  The analysis of the line strengths and significances and the determination 
of plasma emission properties was carried out with Interactive Spectral
Interpretation System (ISIS) tasks. The parameters were determined for 
each of the four spectra, leading to a weighted average of the measurements.
The determination of radial velocities and line widths was carried out 
independently using a variety of {\it ISIS, Sherpa,} and {\it IDL} 
routines, with all approaches leading to consistent results.

\section{Analysis} 

\subsection{ Photometric Time History}                             

\begin{figure}
\vspace*{.1in}
\begin{center}
\includegraphics[angle=90,scale=0.7]{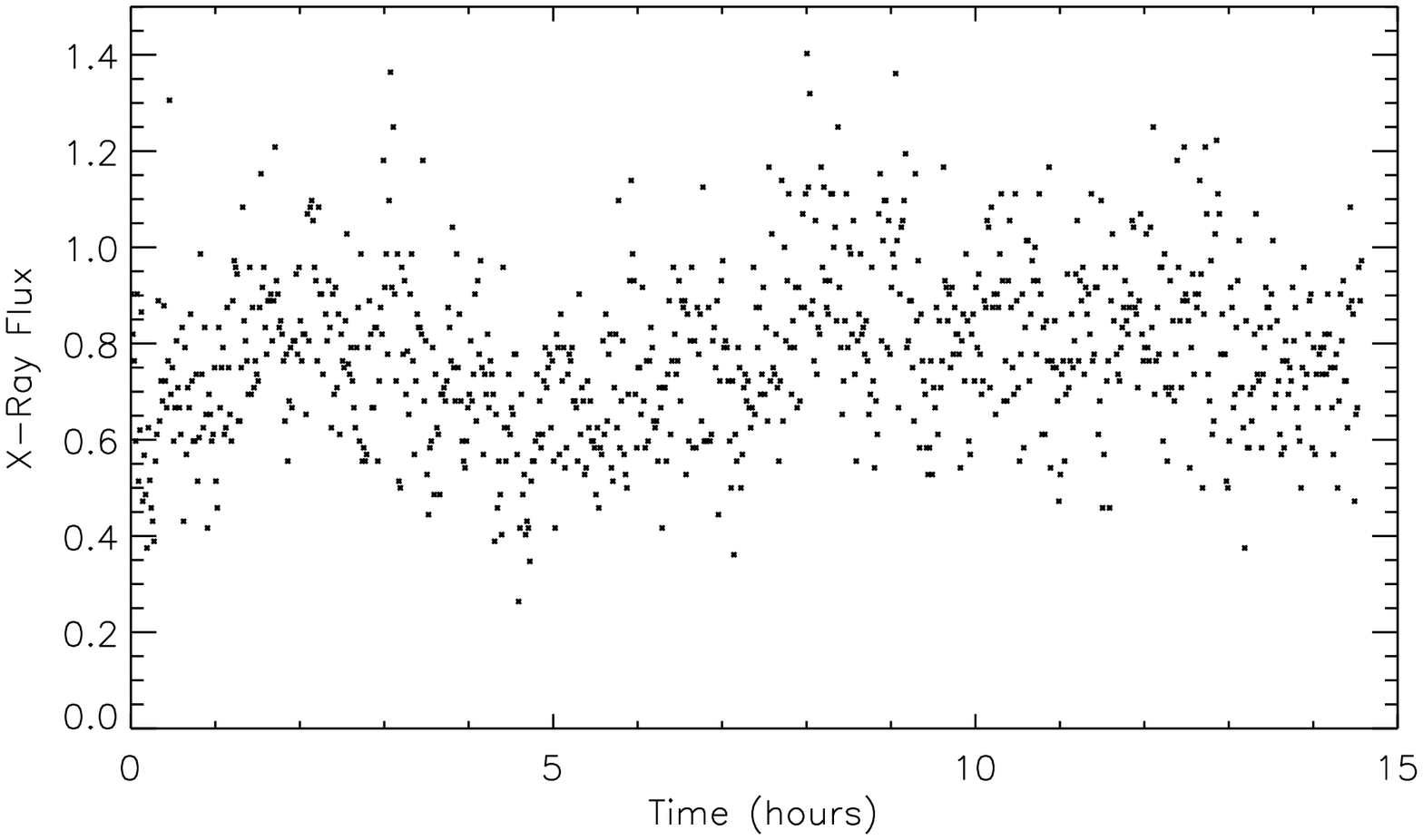}
\end{center}
\caption{Light curve of $\gamma$\,Cas extracted from
our first-order HETGS spectra and binned to 1 minute averages.}
\label{ltcrv}
\end{figure}

  Our observation was conducted at an epoch when $\gamma$\,Cas had just 
gone through a minimum in its $\sim$70 day flux cycle and had an X-ray 
flux close to its average value. 
A light curve during the observation was extracted from HEG and MEG $m$ =
-1 and +1 fluxes. We chose a bin size of 60 s as a compromise between 
effects of photon quantization and the loss of individual shot events.
The resulting light curve is shown in Figure\,\ref{ltcrv}. The plot shows 
a typical meandering pattern of the fluxes with variations on a timescale of 
a few hours. This pattern is typical of a 
few weakly undulating light curves previously seen, such as the 1998 and 
2000 {\it RXTE ``Visit 1"} given by Robinson, Smith, \& Henry (2002; RSH02). 
It is quite different from those in which a slow, probably rotationally 
induced, modulation dominates and clear maxima and minima can be identified. 
Possible dips at  $\le$0, 6, and 14 hours are reminiscent of
the ``7.5 hour cycle"  discovered by Robinson \& Smith (2000).
The cause of this cyclicity (which may not always recur with a cycle 
length of 7.5 hours) is unknown. Nonetheless, this phenomenon
appears robust in most {\it RXTE} data (RSH02).
It was also visible as periodic absorptions in the blue wings of C\,IV 
and Si\,IV resonance lines in archival {\it IUE} spectra (CSR00).

  To see if these flux variations might translate to changes in spectral
line strengths, we 
divided the time sequence into regions of high and low flux and 
generated X-ray spectra from each group. The resulting spectra were 
rather noisy and showed no statistically significant changes. This is 
consistent results presented by SRC98 and Smith and Robinson (1999), who 
showed that RXTE spectra of the shots was generally 
similar to that of the basal component.

  Figure\,\ref{ltcrv} also shows that the lower and upper
envelopes correspond quite well, with the upper envelope being only 
occasionally interrupted by shots lasting 3--4 minutes (e.g., at 3.1 and 
8 hours). Tests with truncated and noised-added renditions of some of
our past {\it RXTE} light curves show that these properties are not unusual. 
Thus, we believe that the envelope correlations and absence of all but a 
few long-lived shots are probably simply artifacts of the low count rate.

\subsection{ General Reconnaissance of Spectrum}

\begin{figure}
\vspace*{.1in}
\begin{center}
\includegraphics[angle=90,scale=0.75]{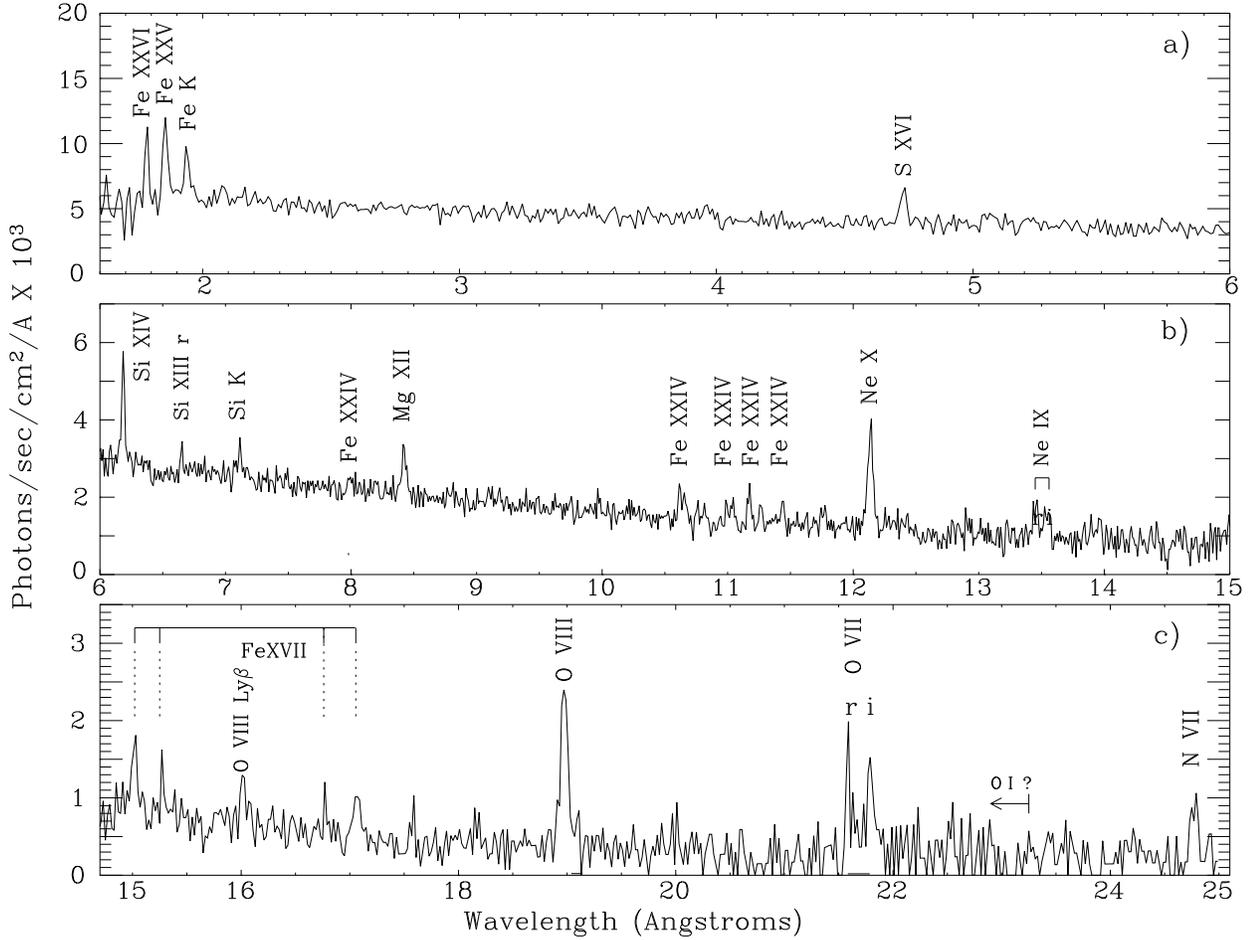}
\end{center}
\caption{The combined HEG--MEG spectrum of $\gamma$\,Cas 
from all four first-order detectors, binned to 10\,m\AA.~ Lines studied 
in this paper are indicated.}
\label{atlas}
\end{figure}

   Figure\,\ref{atlas} exhibits the complete MEG/HEG spectrum of
$\gamma$\,Cas in flux units over the range 1.6\,--\,25 Angstroms,
weighted by the wavelength and detector-dependences in effective
aperture and binned every 10\,m\AA.~ The continuum is strong at the
high-energy end, which in thermal plasmas indicates a hot-plasma,
free-free emission component, and it also exhibits significant
attenuation at wavelengths above 12\,\AA.~ The spectrum is also
sprinkled with a number of emission lines, the properties of which are
listed in Table\,\ref{lines}.  The presence of these lines indicates a
broad distribution of ion stages, ranging from Fe\,$^{25+}$ on the
high side down to N$^{6+}$. The most prominent lines are those of
\ion{Fe}{25} and \ion{Fe}{26} which have been observed by Kubo et
al. (1998). These lines are weak relative to the underlying continuum
or with respect to lines in other $\sim$10\,keV sources.  Most of the
other lines present are Ly\,$\alpha$ lines of H-like even-atomic
number elements, ranging from sulfur to oxygen (but also including
\ion{N}{7}), with one or two Ly\,$\beta$ lines also being present.
Various density indicators are present weakly, including lines of
(He-like) \ion{Si}{13}, \ion{Ne}{9}, and \ion{O}{7}, as well as a
possible weak blend of the \ion{Fe}{17} density-diagnostic at
17.05\,\AA\ and 17.10\,\AA.~ A weak fluorescent K feature of Si at
7.1\,{\AA}, as well as a moderate-strength emission at 1.9\,{\AA},
arising from the analog Fe\,K feature, are also present. These are the
sole visible diagnostics of plasma photoionization processes in the
data.

We ran our spectrum through customized programs as well as ISIS
software to determine the significance of spectral features. The
identification of weak lines of a common ion was guaranteed by their
simultaneous evaluation at predicted wavelengths or in the case of
individual weak features (e.g., \ion{Ne}{10}, Ly\,$\beta$, and
\ion{N}{7} Ly\,$\alpha$) from strengths of lines of various ions along
H-like or He-like electronic sequences, including \ion{Ne}{9} and
\ion{O}{7}.

\subsection{Line Kinematics}

To assess the kinematic properties of the line emitting plasma, we chose 
several strong and unblended lines of H-like Fe, S, Si, Mg, Ne, and O
for detailed study. 
In measuring the radial velocity and nonthermal broadening of the
strongest spectral lines of $\gamma$\,Cas, we found agreement from
two techniques, first, by using appropriate Gaussian-fitting routines 
and, second, by performing a cross-correlation and convolution of 
the profiles of these lines in an archival spectrum of AB\,Dor with
respect to our $\gamma$\,Cas data. The resulting centroid wavelengths, 
standard deviation, Gaussian-fitted line center and continuum fluxes 
are listed in Table \ref{lines}. The measured wavelengths indicate 
no systematic radial velocity for the emission region. 
The line widths are all consistent with a Gaussian velocity
distribution of $\sigma_v=478\pm50$ km\,s$^{-1}$. The error bars for
the \ion{Fe}{25} and \small{XXVI} lines are considerably larger, and
we estimate these to be $\pm325$ km\,s$^{-1}$.  No statistically
robust evidence was found for asymmetries in the line
profiles. However, because there is a hint of stronger wings than
indicated by Gaussian fits among the stronger Ly$\alpha$ lines, we
cannot rule out the possibility of a second kinematic component with a
larger broadening velocity.

To determine the fluxes of weaker lines, such as those of iron L-shell ions,
we perform similar Gaussian fits fixing their wavelengths at the known
theoretical values and the velocity broadening obtained from the fit to strong
lines mentioned above. The line fluxes and continuum level at the line centers
are also listed in Table \ref{lines}. The equivalent widths in \AA\ can be
computed by dividing the line fluxes by the continuum fluxes.

\subsection{Global Spectral Analysis}
\label{globl} 

\subsubsection{The basic model ``M1"}

  In order to derive the thermal properties of the plasma, we chose to
use global fitting techniques to constrain the emission measure
distribution of the plasma. We utilized the Astrophysical Plasma
Emission Code (APEC), as implemented in ISIS, to perform the spectral
fitting.  From previous X-ray satellite observations of $\gamma$\,Cas,
we know that the continuum emission can be described by a heavily
absorbed bremsstrahlung with a temperature $\simeq10$\,keV. The high
spectral resolution of the \textit{HETGS} clearly demonstrates the
existence of more complex emission measure structure at lower
temperatures, although the overall continuum seems to be dominated by
the $\simeq10$\,keV component. Within the {\it HETGS} bandpass we
found that the temperature and the absorption column density are
somewhat degenerate. Therefore, we constrained our spectral analysis
to contain a hot component with a fixed temperature of 12.3 keV, as
determined by the broad-band {\it BeppoSax} observations (Owens et al.
1999). We consider the errors on this parameter to be about
$\pm{1}$\,keV. As Table\,\ref{fekt} shows, the value of 12.3\,keV is
generally consistent with the previously reported temperature
determinations.
 
The presence of relatively strong \ion{Fe}{23} and \small{XXIV} lines,
the ratio of the \ion{Fe}{17} 15 {\AA} to 17 {\AA} lines (cf. e.g.,
Kinkhabwala et al. 2002), as well as the values of the He-like triplet
G-ratios (($i+f$)/$r$) of \ion{Ne}{9} and \ion{O}{7} are all
consistent with collisionally ionized plasma. These ratios are also
consistent with a photoionized plasma if the photoexcitations
contribute significantly to the resonance lines (Kinkhabwala
et~al. 2002). However, the latter condition does not seem to apply for
the observed spectrum, since this would also require much larger
high-$n$ series line fluxes for H-like and He-like ions populated by
photoexcitation, in contradiction with the almost non-existent
He-$\beta$ line of \ion{O}{7} (18.6\,{\AA}) and the only weak
Ly\,$\beta$ line of \ion{O}{8} (16.0\,\AA) in the data.  The absence
of detectable radiative recombination continuum features (e.g., for
\ion{S}{16} at 3.5\,{\AA}, for \ion{Si}{14} 4.6\,{\AA}, for
\ion{Mg}{12} 6.4\,{\AA}, for \ion{Ne}{10} 9.3\,{\AA}, \ion{O}{8} at
14.2\,{\AA}) is also inconsistent with photoionization
models. Therefore, we will pursue only the collisional models to
explain the hot plasma.

\begin{figure}
\vspace*{.1in}
\begin{center}
\includegraphics[angle=0,scale=0.75]{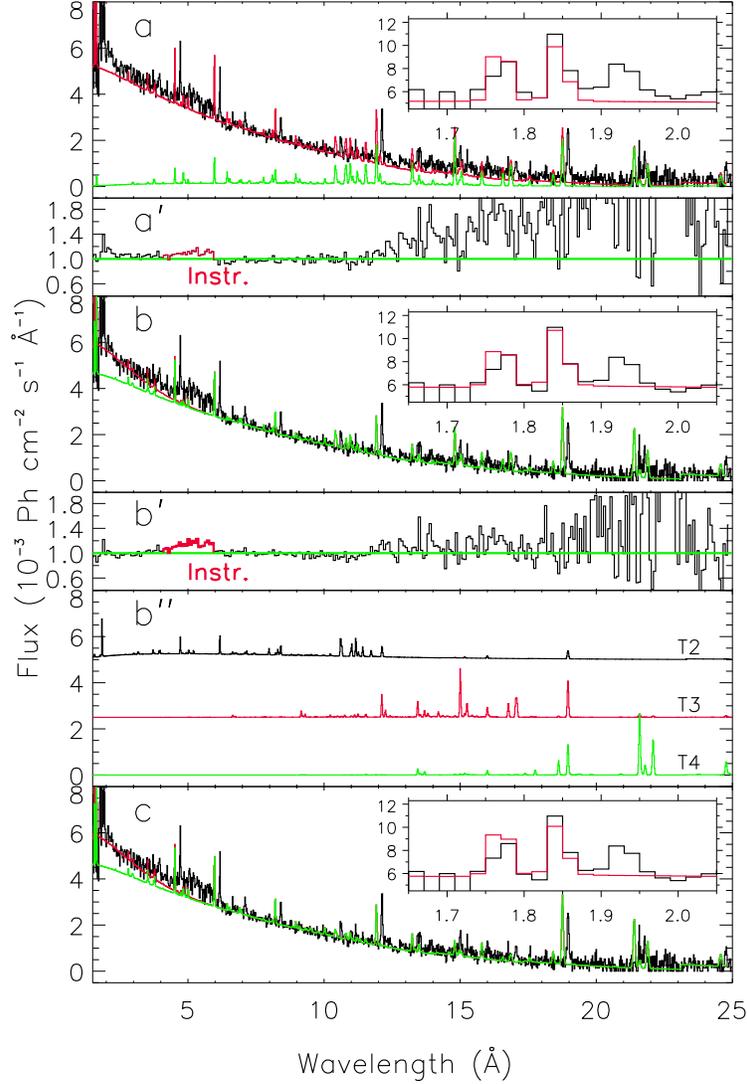}
\end{center}
\caption{Comparison of best fit models and the data. Panels (a), (b),
\& (c) correspond to models M1, M2, \& M3, respectively. In all
panels, the black line depicts the data, red is the best fit model,
shifted by -0.2\,\AA~ for clarity. In panel (a), green is the
contribution from the components $T_{\rm 2}$, $T_{\rm 3}$, and $T_{\rm
4}$, also shifted by -0.2\,\AA.~ In panels (b) and (c), red is the sum
of all components while green is the sum from all components ($T_{\rm
1-4}$) having the ``low" column density of $\approx3\times10^{21}$
cm$^{-2}$.  Panels (a$^{\prime}$) and (b$^{\prime}$) depict the
data-to-model ratio for models M1 and M2.  Panel (b$^{\prime\prime}$)
shows the individual spectra (unshifted) for components $T_{\rm 2}$,
$T_{\rm 3}$, and $T_{\rm 4}$ of Model 2.  The insets show the \ion{Fe}{26}
and \small{XXV} line region in detail; the obvious mismatch of the
fluorescence feature at 1.9\,\AA\ is discussed in the text.  Note that
the 4-6\,\AA\ region is not well calibrated due to an iridium spectral
edge from the telescope mirror coating.}
\label{demmod}
\end{figure}

From the presence of lines from both Fe L-shell and
\ion{O}{7}--\small{VIII} ions, it is clear that emission regions also
exist with temperatures from a few keV down to $\sim$100 eV. As a
first attempt to characterize the differential emission measure (DEM)
in this regime, we chose a four-component thermal and optically thin
model, which we call M1, to explain the main characteristics of the
line and continuum emission. Each temperature is restricted to a
certain range to represent the characteristic emission lines. Although
the temperature of the dominant hot component was fixed at k{\it
T$_{1}$} = 12.3\,keV, the value of the second component was permitted
to float within a range of 1 and 4\,keV. This condition was imposed by
the presence of \ion{Fe}{23} and \small{XXIV} lines at 10 to
11\,{\AA}. The third component has a temperature between 300\,eV and
1\,keV, as suggested by the relatively strong \ion{Fe}{17},
\ion{Ne}{9}, and \small{X} lines. The fourth component has a
temperature near 100\,eV and accounts for the \ion{O}{7} and
\ion{N}{7} lines. We will refer to components 2--4 as ``warm'' plasmas
in our discussion below.  To summarize, the requirements for 4
components are determined by lines arising from ions having ionization
potentials in different ranges, which in turn correlate very well with
wavelength. Thus, we may say that component 1 is determined by the
hard continuum and ratio of the two Fe\,K lines ($<$1.8\,{\AA}, and
components 2, 3, and 4 by lines in the ranges 1.8\,--\,11.9\,{\AA},
12\,--20\,{\AA}, and $>$20\,{\AA}, respectively. This is illustrated
for our ``Model 2" in Fig\,\ref{demmod}b$^{\prime\prime}$, which are
both discussed below.

Previous satellite missions have found that the measured \ion{Fe}{25}
and \small{XXVI} line intensities require a significantly sub-solar Fe
abundance (see Table\,\ref{fekt}).  Therefore, we decoupled the Fe
abundance in the hot component from the abundances of all metals
(including Fe) in the warm components.  The measured line strengths
probably have errors of $\pm{10}$\,--\,15\% and largely determine the
uncertainty of the Fe abundance for the hot component, which we take
as $\pm{20}$\%.  The uncertainty in the abundance estimates of the
other elements is sensitive to the details of DEM models, and may be
even larger. In model M1 all four components are affected by a common
absorption column density, and have the same velocity broadening with
the width fixed at the value derived in the previous section. In our
models each component has three unknowns (temperature, abundance, and
column density). Since in model M1 we tie together the column
densities and separate Fe from other metallic abundances in the hot
component, we end up formally with 10 free parameters.  A joint fit to
the $\pm 1$ orders of HEG and MEG spectra was carried out with ISIS,
making use of Cash statistics to take into account the small number of
counts in the long wavelength region.

The best fit parameters for model M1 are shown in Table\,\ref{params}.
Because we are concerned only with the overall temperature structure 
of the plasma, and we do not expect the simple models discussed here to 
be statistically acceptable across the entire HEGS bandpass, we have 
not attempted to derive statistical uncertainties of all the
best fit model parameters. 
Formal errors that one could calculate from line and continuum 
discreprancies, which are themselves highly interdependent, 
do not reflect the probably larger errors inherent in the qualitative
representation of the geometry envisaged for the X-ray sites near
$\gamma$\,Cas. The input geometries can be quite diverse, depending for 
example, on the proximity to and association of the X-ray emission centers 
with the Be disk and also their distribution over the Be star's surface. 
In addition to geometrical considerations, the emission  
mechanisms associated with a supposed degenerate companion could be even more 
different and thus more elusive to quantify. To give the reader a flavor
of the likely range of parameters within the context of the Be star-disk 
model of X-ray emission, we report results on three models.
We represent the errors for most parameters in terms of the measured 
difference between the M1 and the M2 model discussed below. We also exhibit 
the comparison between each model and data in Figure\,\ref{demmod}a 
graphically. For illustration purposes, the $\pm 1$ orders of HEG and 
MEG spectra are rebinned to 0.02\,{\AA} binsize and co-added.

  As expected, the emission measure and spectrum below $\simeq$\,6\,\AA\ are 
dominated by the hot component. Note from the insets in Figure\,\ref{demmod} 
that the Fe line ratio for the hot component fits the observations reasonably
well, strengthening the conclusion that the dominant plasma emission 
processes are thermal. Note also the clear presence of the Fe K-fluoresnce
feature, which is not included in the APEC model. 
In our first model, M1, we find an Fe
abundance of only 0.34 for the hot component, in agreement with the
findings of previous X-ray missions (Table\,\ref{fekt}).
The abundances of the metallic elements in the warm components are
constrained by the strengths of the Ly$\alpha$ lines of H-like and He-like 
ions. The overall good agreement between 
the solar abundance model and data for these lines indicates that the 
metallic abundances of the warm component are significantly closer to 
solar values than Fe is in the hot component. 

   One obvious mismatch in model M1 is the He-like triplet
line ratios of O and Ne, and our discussion of these is deferred to
$\S$\ref{herat}.
Besides the continuum discrepancy in the 4--6\,{\AA} region, which may well 
be attributed to the known calibration problem in the effective area of
\textit{Chandra} mirror assembly, two additional spectral regions be can be
easily identified from Figure\,\ref{demmod}a where the model continuum level 
is significantly lower than the data. The first is the region between 2 and
2.5\,{\AA}, and the second is at $\gtrsim 12$\,{\AA}. The green line in
Figure\,\ref{demmod}a shows the contribution from the 
three warm components.
It is clear that the continuum flux from this component is too small to account 
for the deficit. The shallower slope of the observed continuum near 12\,{\AA}
indicates that the absorption column density may be lower than what is
derived from model M1. Yet, a lower column density would worsen the
discrepancy in the 2--3\,{\AA} region where the observed continuum slope is
steeper than the model.

\subsubsection{The two-column density models (hot component)}

\begin{figure}
\vspace*{.1in}
\begin{center}
\includegraphics[angle=0,scale=1.0]{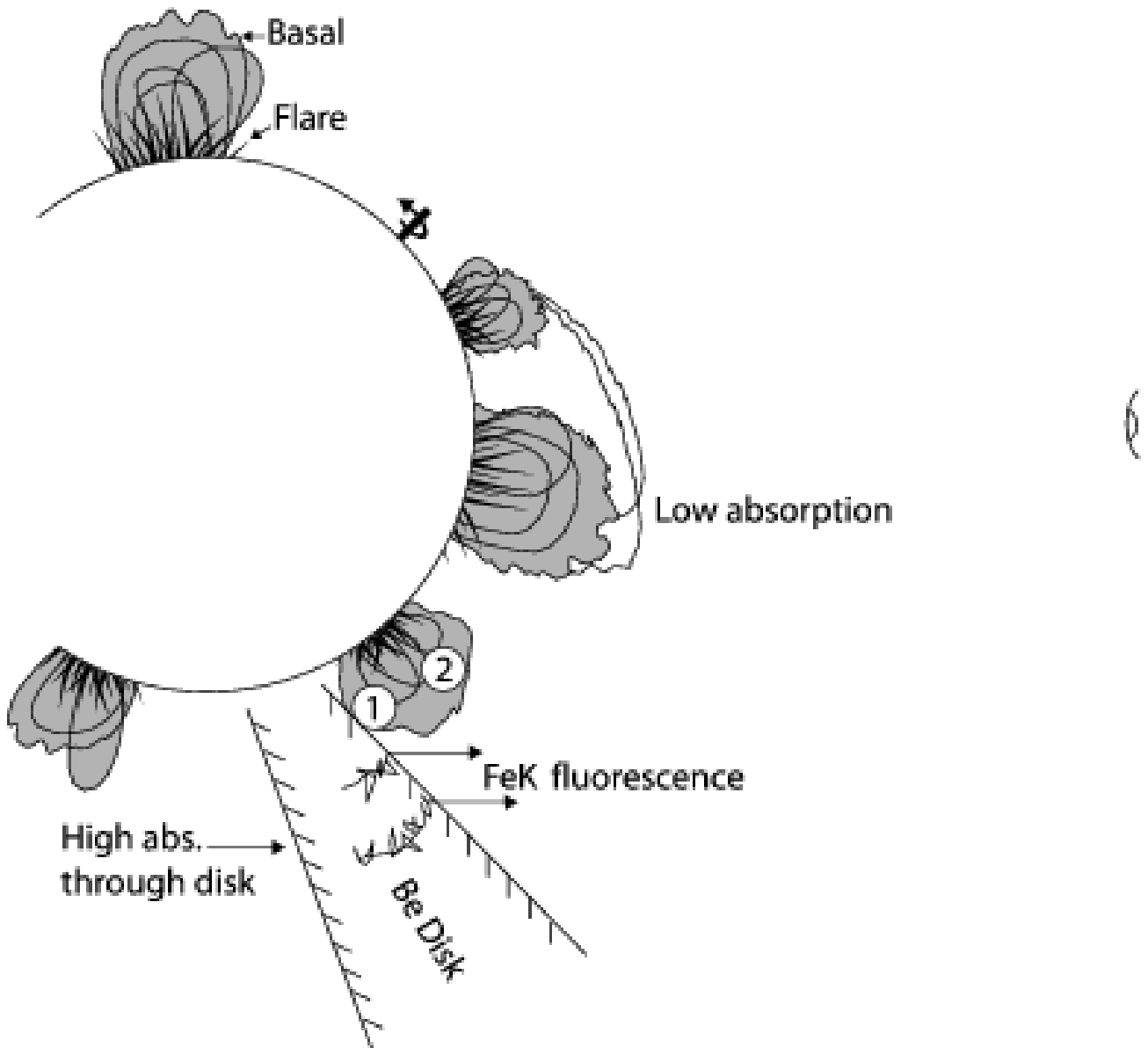}
\end{center}
\caption{Sketch of a cross-section of $\gamma$\,Cas inclined at 45$^o$ to the
observer at right. In our picture (see also SRC98) the star's surface
is populated by small X-ray flare centers, which fill canopies within 
magnetic loops as they explode. The observed hard X-rays are produced
in both regions. The Be (decretion) disk is in the star's equatorial plane 
and absorbs all of the soft X-rays but rather little of the hard X-rays,
and a fraction of the hard flux emitted from the canopies viewed behind
the disk in the lower left of the figure.
The disk is also 
the presumed site of Fe, Si K fluorescence photons, which are end 
products of photoionization from very hard continuum X-ray photons. 
Symbols `1' and `2' correspond to sites in a typical basal 
flux-emitting canopy where soft-X-rays are generated either from canopy-cloud 
collisions with the disk {\it (case 1)} or cooling within the canopies {\it (case 2).}}
\label{cartoon}
\end{figure}

  One may avoid this problem by varying the column density absorptions
of the dominant emitting components. This can be done in one of two
ways.  In the first case, one can increase the column density for the
hot component and decrease the column densities of the warm ones to
compensate the flux attenuation where needed. This procedure required
only a slight increase in the column density of the hot component but
resulted in a decrease of the warm component column to effectively
zero.  However, these changes slightly reduced the continuum flux
below 10\,\AA\ and raised the fluxes at long wavelengths. Even so, the
corrections were not sufficient and still generated an undercorrection
in the 12--16\,\AA\ region.  We were able to produce a better fit by
splitting the hot component into two subcomponents, one with a rather
high column density of 10$^{23}$ atoms\,cm$^{-2}$ (a value to be
justified later), and a second one with an lower, floating column
density, which is tied to the attenuation of the warm components. Note
that a column density of 10$^{23}$ atoms\,cm$^{-2}$ transmits nearly
all short wavelength ($\le$ 4\,\AA)~ flux while at the same time
effectively absorbing flux above 12\,{\AA}.  We allowed the volume
ratio of the more absorbed to less absorbed hot subcomponents to vary
freely, and this value floated to 15\%.  However, we found that
fractions as high as 30\% or as low as 10\% also give acceptable
fits. Based on these results, we modified the model M1 to produce a
model M2 with a fraction of 25\% for the high-column density
subcomponent and 75\% for the low-column subcomponent (this causes
minor adjustments in some of the other fit parameters).  This is
exhibited as a sketch in Figure\,\ref{cartoon}.  The best-fit
parameters are listed in Table\,\ref{params}, and the comparison
between the model and data are shown in Figure\,\ref{demmod}b.  This
model gives a significantly better fit to the data than M1.  The
extra, more heavily absorbed hot component not only fills in the
continuum deficit near 2\,{\AA}, but it also allows the other plasma
components to take on a slightly lower column density, $2.7\times
10^{21}$\,cm$^{-2}$, thereby removing the disagreement in the longer
wavelength regions. The various predicted line emissions appear to
agree with the data satisfactorily as well. We note that in M2 the Fe
abundance is decreased to 0.22, mainly from the \ion{Fe}{25} and
\small{XXVI} lines. We note that the model fluxes are slightly high
for the the \ion{Fe}{17} lines. However, such small discrepancies can be
easily attributed to our oversimplified DEM models.

In the models M1 and M2, we have tied the iron abundance in the warm
components to the other metallic elements. It is interesting to
investigate if it is possible that the iron abundance in both hot and
warm components are the same, and very sub-solar. For this purpose, we
constructed a model M3, which is identical to M2, except that the iron
abundance of the warm component is tied to the value determined for
the hot one (now 0.26), and it is independent of other metallic
abundances.  Of these three models, M2 gives the best overall fit, and
thus we adopt it.  The best-fit parameters for all three models are
listed in Table\,\ref{params} and the comparisons between the models
and data are shown in Figure\,\ref{demmod}c. It is evident that for M3
the Fe L-shell lines, especially \ion{Fe}{23} and \small{XXIV} lines,
are underpredicted from the low trial iron abundance. This low Fe
value is required by the \ion{Fe}{25} and \small{XXVI} lines in our data and by
observations of other satellites (see Table\,\ref{fekt}).  Therefore,
we conclude that the iron abundance in the warm component is close or
equal to the solar value.

The difference in Fe abundance of the hot and warm components argues
strongly that these plasmas are not cospatial. Additionally, the hot
component must be fairly sharply peaked at $\ge$10\,keV or it would
overpredict the \ion{Fe}{23} and \small{XXIV} line strengths, even
with the low Fe abundance. It is not clear whether the first two warm
components are distinct from each other, or if the third warm
component is distinct from these other two.  We have also experimented
with a power-law DEM component for the temperature range 0.5--3\,keV,
which fits the data as well as M2. The quality of data does not enable
us to differentiate a smooth temperature distribution or several
discrete temperatures for the warm components. However, a single
power-law DEM model that includes the hot component fails to describe
both continuum and line emission satisfactorily. Thus for simplicity
sake, we have adopted the multi-temperature thermal models as the
preferred analysis method.

It is worth mentioning that a strong neutral oxygen absorption edge at
23.05\,{\AA} is predicted in all of our models. Because the continuum
level at this wavelengths is very low, and the HETGS effective area
small, the statistical quality of the data does not allow for a
straightforward analysis of this feature. Nevertheless, we may still
carry out a statistical test to see whether this absorption edge is
present in the data by extracting two small spectral windows
immediately below and above the edge at 23.3--23.8\,{\AA} and
22.4--22.9\,{\AA}. There are no known strong line emissions in these
two wavelength intervals.  The flux ratio of these two windows in the
HETGS data is F(23.3--23.8)/F(22.4--22.9)= 0.94 $\pm{0.22}$, while the
corresponding ratio is higher, equaling 2.8, for model M2. However,
since the quoted uncertainties account only for finite photon
statistics, we feel it is premature to rule out the possible existence
of the \ion{O}{1} edge.  We are planning to search for this feature in
upcoming \textit{XMM-Newton} observations.
  
To summarize this analysis, although it is hard to quantify the exact
uncertainties on each parameter in a complex thermal model, we have
found: (a) two separate column densities are required for the hot
component, (b) several warm temperature components, or a continuous
temperature distribution of warm plasma with several peaks, is
required in addition to the hot component, and (c) the hot component
has a significantly lower Fe abundance than the warm components.
These conclusions are robust, as indicated by the inability of
alternative models to provide good global fits to the data.

\subsection{Comparison of X-Ray and Ultraviolet-Derived Column Densities }
\label{colden}

 The analysis in the previous section established that both the hot
and warm X-ray components are absorbed by a gas with a column density
of 3$\times10^{21}$ atoms\,cm$^{-2}$. This value, which is again
partly a consequence of our assuming a temperature of 11--12\,keV from
the literature, is consistent with the results of previous X-ray
studies, which find column densities in the range of
$10^{21}$\,--\,$10^{22}$ atoms cm$^{-2}$ toward $\gamma$\,Cas (see
Table\,\ref{fekt}). Most of these in previous studies attribute the
absorption to the interstellar medium (ISM).  For the {\it RXTE} and
{\it Tenma} studies, the instrumental flux calibrations below
2--3\,keV were not as well determined, so the rather high column
density estimates derived in these cases are suspect.  The remaining
estimates cluster tightly at 1.5--1.6\,$\times$10$^{21}$ atoms
cm$^{-2}$. In contrast to the X-ray results, the column densities
derived from UV studies of ISM resonance and Lyman$\alpha$ lines show
much lower values.  In the case of $\gamma$\,Cas, the broad absorption
core of Lyman\,$\alpha$ is produced by an ISM column density of
1.4$\times$10$^{20}$ atoms cm$^{-2}$ in {\it Copernicus} spectra
(Bohlin, Savage, \& Drake 1978). Various {\it IUE} studies (e.g., Van
Steenberg \& Shull 1985, 1988, 2$\times$10$^{20}$ atoms cm$^{-2}$;
Diplas \& Savage 1994, 1.5$\times$10$^{20}$ atoms cm$^{-2}$) show an
excellent consistency. Moreover, the resonance line results for five
metallic ions, relative to the expected depletions in the ISM of
metal, also bear out the N$_{H}$ result determined from the Lyman
$\alpha$ line core measurement (Van Steenberg \& Shull 1988).  To
summarize, the UV metal and hydrogen lines towards $\gamma$\,Cas are
all low and show {\it inter alia} agreement. In contrast, X-ray
spectra lead to column densities some ten times higher than the UV
results, according to the wavelength attenuation of the soft X-ray
continuum. The excess absorption in the X-ray case (the 10$^{21}$
atoms cm$^{-2}$ component) can be easily explained by their being much
more cold gas near the star than ISM material along the line of sight
to it.

\subsection{Analysis of Diagnostics of Volume Density}
\label{herat} 

Our spectrum reveals density diagnostics, including partial sets of
three ``$rif$" triplets of He-like \ion{Si}{13} (6.6--6.7\,\AA),
\ion{Ne}{9} (13.4--13.5\,\AA), and \ion{O}{7} (21.6--22.1\,\AA) as
well as the 17.05\AA, 17.10\,\AA\ lines from the strong $2p^{6}$
$\rightarrow$ $2p^{5}3s$ transitions of Fe\,XVII. The forbidden $f$
component of \ion{O}{7} is too weak to be detected in our spectrum.
Likewise, for \ion{Si}{13}, the intercombinational ($i$) component is
not visible, and the ($f$) component component detection is
doubtful. The latter components lie on the Si edge of the detector
responses, so the I($f$)/I($i$) ratio is indefinite and can add no
information. For \ion{Ne}{9} and \ion{O}{7} it is more apparent that
the ratio (I(i) + I($f$)/I(r)) $\leq$ 1 and that I($f$)/I($i$)
$\simeq$0. (The former ratio is visible for \ion{Si}{13} as well.)
The first ratio argues that the plasma is collisionally dominated
while the second indicates that the forbidden component is quenched
either by collisions or photoexcitations. The collisional
interpretation for the $f/i$ ratio would set the plasma density at
$\sim$3$\times10^{11}$ cm$^{-3}$ and $\ge$3$\times10^{10}$\,cm$^{-3}$
for the \ion{Ne}{9} and \ion{O}{7} ions, respectively.  One can
perhaps discriminate between the collisional and photoexcitation
options for $\gamma$\,Cas by comparing the photoexcitation and decay
rates of the 2\,$^{3}$S $\rightarrow$ 2\,$^{3}$P transition
appropriate to a 1000 \AA\ radiation field of a B0.5\,Ve star. Scaling
from rates given by Kahn et al. (2001), one finds that radiative
dominance requires only that the Be star be nearer to the irradiated
plasma than 10R$_{*}$ and 40R$_{*}$, respectively. In the picture of
X-rays being produced by an accreting white dwarf, the corresponding
limits would mean a similar dilution factor of
10$^{-2}$\,--\,10$^{-3}$. In the Be X-ray paradigm, the hot sources
are no further from the star than the magnetic co-rotation radius,
which is probably less than one stellar radius.  Since the
photoexcitation diagnostics turn out to be only weak constraints, and
because both the density and irradiation routes are available to
quench the forbidden components, little new information is actually
gained from the $f/i$ ratios.

A final estimate of the densities can be obtained from the
dimensionless ionization parameter $\Xi$ = $L_{{\rm x}}/(4\pi
r^{2}\,c_{\rm s}\,P)$ (e.g., Nayakshin, Kazanas, \& Kallman 2000) and the
fact that collisional processes dominate whenever $\Xi < 1$. Here
$L_{\rm x}$ is the luminosity of the hot X-ray flux, $r$ the distance to
gas particles illuminated by these sources, $P$ the local gas
pressure, and $c_s$ the speed of sound.  Thus, a lower limit on the
gas density is obtained by assuming $\Xi$= 1 and solving for the
pressure, which can be converted to density by selecting an
appropriate temperature. As an example, we consider the situation in
which hot X-ray sources strongly illuminate a nearby reservoir of cold
gas, such as the Be disk. We assume that the disk is at least as far
away from the illuminating sources as the size of the hot X-ray
regions themselves ($\sim10^{11}$\,cm; estimated by SRC98). We further
assume a temperature of 10$^6$\,K for the average warm component.
From these values, we obtain the limit $n_{\rm H}$ $>$ 10$^{9}$
cm$^{-3}$ for the gas density of the warm component plasma.

\vspace*{-.15in}

\section {Discussion  }

  In this section we acknowledge that while an X-ray spectrum can provide
a glimpse of the physical conditions in the hot plasma sites where the 
flux is produced, it is not likely to point the way to a truly unique model 
or geometry of these regions. Nonetheless, 
we do believe there are tell-tale hints
of their proximity to the Be star. In this section we will first examine the 
Be star model, paying particular attention to our column density estimates 
and to the role of cold gas in providing fluorescence and attenuation of 
the spectrum. The influence and hence proximity of this material can 
provide evidence pertinent to this picture that cannot be obtained from 
other kinds of observations. We then compare the X-ray spectral properties
of $\gamma$ Cas with those of known binary systems containing an active
white dwarf companion.

\vspace*{-.15in}

\subsection{Relationships among the X-Ray Emitting Components }

\subsubsection{Fe fluorescence: ``cold matter" }

  Our measured equivalent widths of the Fe and Si K fluorescence features 
are -19\,m\AA\ and -5\,\AA,~ respectively. 
The presence of these lines indicates that substantial ``cold" gas is
present along the lines of sight to the hot X-ray emission sites.
Although modeling the Fe fluorescence feature is quite complicated and
depends on details of the temperature of disk particles and disk geometry, 
we can get a rough estimate of the formation parameters for this line by
adopting a simple model, which is sketched in Figure~\ref{cartoon}. 
Consider the fluorescence within a cold, plane-parallel slab illuminated 
by numerous hard X-ray sources placed just behind and 
in front of it. If the slab is optically thin to X-ray photons with 
wavelengths below 1.94\, \AA,~ some of the photons will photoionize the 
K shell of Fe atoms, and some 1/3 of them will emit 
fluorescence photons with wavelengths near 1.94\,\AA\ (Liedahl 1998).~ 
Because fluorescence is an isotropic process, it will produce essentially 
the same feature whether the optically thin slab is illuminated from
the front or back by the hard X-ray sources. Now stipulate further 
that the slab has a solar composition and an arbitrary column thickness 
of 10$^{23}$ atoms cm$^{-2}$. Its X-ray spectrum should also be flat in 
the 1.3\,--\,1.94 \AA\ region regions of the Fe K continua. Under these 
assumptions Kallman (1991) has computed an equivalent width of -69\,m\AA\ 
for the 1.94\,\AA\ fluorescence feature. The geometry just described 
provides a reasonable approximation of the X-ray sites and disk thought 
to be associated with $\gamma$\,Cas. 
The value of 10$^{23}$ atoms cm$^{-2}$ we have used here and elsewhere in 
this paper is taken from the predicted (vertical) disk column density 
of Millar and Marlborough (1998) for the $\gamma$\,Cas disk. Moreover,
we have already found that it is a reasonable estimate of the column
density according to the long-wavelength attenuation in the HETGS spectrum.   
In the general picture we have described, hot X-ray sites 
are evenly distributed over the surface of the Be star and 
illuminate the Be disk. Fe continuum photons from these
sites are converted by Fe atoms in the Be disk to $\approx$1.94\,\AA\ 
photons by the fluorescence process and are emitted isotropically. 
Since the sites reside near the
star, no more than half the outward directed flux can illuminate the disk.
Thus, in this rough approximation, we might expect the equivalent width 
of the Fe fluorescence line to be about -69/2 = -35\,m\AA.~ 
This figure agrees reasonably well with the observed value of -19\,m\AA.~ 
The discreprancy of a factor of two in this estimate might be attributable 
to the actual decreasing continuum slope shortward of the K feature or 
to an oversimplified distribution of the hot plasma sources. 
Considering the uncertainties in our picture, we regard the observed 
strength of the fluorescence feature to be consistent with the presence 
of intervening disk particles having the column density estimated by 
Millar \& Marlborough (1998; MM98).

\subsubsection{Origin of the two absorption systems }

  The two-column absorption model required for the hot emission component
in $\S$\ref{globl} is an expected outcome if the emission sites
themselves are sprinkled closely around the Be star, such that some
fraction of them are behind the disk, but still in our line of sight.
Various correlations of X-ray fluxes with UV diagnostics 
indicate that the flare and basal components are probably formed at or
at most within a few tenths of a radius of the star's surface 
(see SRC98, SRH98, Smith \& Robinson 1999). Referring to 
the sketch in Figure~\ref{cartoon}, if the X-ray active sites are 
distributed uniformly over the star and close to the surface 
the ratio of the sites in front
of the Be disk to those behind it will be $\simeq$0.25.
We used this geometrical visualization in $\S$\ref{globl} as well as the
MM98 disk thickness for the ``high" column density to arrive at the ratio 
of two absorption columns of the hot subcomponents for our models M2 and M3. 

There are two possible CS gas structures near the Be star that could
be responsible for the ``low" column density (3$\times10^{21}$
atoms\,cm$^{-2}$) required from our modeling of the soft X-ray
continuum. The first is any intermediate-latitude gas residing near
the top and bottom boundaries of the Be disk, while the second is the
intermediate latitude wind. In the first instance, soft X-rays can be
created and partially absorbed within the outer boundaries of the
disk. However, this requires a fine-tuning of the placement of the
emission volumes within the upper and lower boundaries of the disk,
which we deem unsatisfactory. The possibility of wind absorption seems
easier to accept. For example, Cranmer, Smith, \& Robinson (2000) have
discussed the wind column density of $\gamma$\,Cas lies in the range
$10^{21}$\,--\,$10^{22}$ cm$^{-2}$. Since this agrees very well with
our result for $N_{\rm H}$ in Table\,\ref{params}, the wind can be
regarded as the leading contender for the source of the low-column
component of X-ray absorption.

\subsubsection{The warm X-ray emitting components}\label{warmcool}

The rather different Fe abundances of the hot and warm plasmas, as
well as the likely break in the DEM between these temperatures,
indicates that the corresponding emission volumes are not
cospatial. The combination of a 12\,keV component with one or more
components with temperatures extending down to 0.1 keV is very unusual
in a B star.  This leads to the question of whether the hot and warm
component volumes are in close proximity and indeed might be causally
related.  It is possible, for example, that the hot and warm
components are different manifestations of the same driving process,
such as the the magnetic field reconnections which are thought to
create the flare-like shots.

Alternatively, we note that the hot plasma has 7--10 times the
emission measure of the warm plasma and also a higher energy per
particle.  Since the thermal energy within hot plasma is as much as 50
times the energy within the warm X-ray emitting regions, it is
feasible energetically that the warm plasma is heated by processes
occurring within the hot plasma regions. An obvious possibility is
that warm component arises from the photoionization of nearby cold gas
by the hot X-ray flux. On the other hand, we have already all but
ruled out this possibility because, except for the fluorescence
features, the prevailing diagnostics in the spectrum clearly favor a
collisional source of heating.  Such heating can arise in a number of
ways. The first, {\it case 1,} is that {\it the hot component heats
pre-existing gas disk by colliding with it.} To see how this might
happen, recall that SRC98 found from plasma cooling arguments that the
heated plasma from initial flaring on the stellar surface will expand
rapidly and without much energy loss, so that it is free to collide
with preexisting stationary gas in, for example, the Be disk.  A
second possibility, {\it case 2}, is that the warm components are
simply {\it the result of radiative cooling of the hot plasma.}
Cooling curves of hot plasma (e.g., Cox 2000) show the presence of
plateaus or minima in the cooling rate at temperatures of $\simeq$
0.1\,keV, 0.3-0.4\,keV, and 2\,-\,3\,keV. These are values very
similar to those of components \#2\,-\,4 in our models (and also
temperatures found for the B0.2\,V star $\tau$\,Sco), and suggest that
the triple-peaked DEM of these components could be a consequence of
plasma accumulating at these temperatures as it cools from its initial
high value of $>10$ keV.  However, this picture presents a difficulty
of explaining how the Fe abundance in the very same plasma could
increase to the solar value as it cools. We will return to this
question below.

{\it Case 2} suggests that the warm component plasma resides in
co-rotating volumes attached to the star's surface (see SRC98, SRH99)
while {\it case 1} favors the inner regions of the Be disk, which have
a Keplerian rate. In the range of interest, within one stellar radius
of the surface, the two velocities are rather similar. Consider that
the rotational velocity $v\sin i$ of $\gamma$\,Cas has been variously
measured in the range 230\,-\,380 km\,s$^{-1}$ (Slettebak 1982,
Harmanec 2002), and larger velocity broadenings cannot reasonably
arise from either the photosphere or from a disk in Keplerian
orbit. Then the observed X-ray line broadening of 478\,km\,s$^{-1}$ is
consistent with material forced into co-rotation at fairly low
altitudes over the Be star's surface, ({\it case 2}). However, this
broadening is also consistent with the orbital velocity of the inner
disk, the environment we associate with {\it case 1}, for which
broadening of high-level hydrogen lines has been found up to 550
km\,s$^{-1}$ (Hony et al.  2000). Thus, the broadening of the X-ray
lines cannot be use to discriminate between the cooling residue and
colliding ejecta pictures.

\noindent {\it Analogs in other B stars ($\tau$\,Scorpii):} 

We can also consider alternate causes for the production of
warm-component X-rays in an ostensibly single, early B-type main
sequence star. Possibly the best such case is the B0.2\,V spectral
standard, $\tau$\,Sco.  The UV spectrum of this star is anomalous in
that it exhibits P\,Cygni profiles in its resonance \ion{O}{6},
\ion{N}{5}, and \ion{C}{4} lines. The absorption components of these
profiles are abnormally broad while the emission components are
significantly redshifted. Analyses of {\it ASCA,} {\it
Chandra/HETGS,}, and {\it XMM/RGS} spectra (Cohen, Cassinelli, \&
Waldron 1997, Cohen et al. 2003, Mewe et al. 2003) demonstrate that
the X-ray emissions have a complicated emission measure that extends
to high temperatures. The high-resolution studies showed that the
X-ray emission measure is composed of as many as three components with
k{\it T} values centered near $\ge$2.5, 0.7, and 0.1\,keV. According
to predictions of standard wind theory, the high temperatures
corresponding to first two components are unlikely to be attained by
wind shocks.  However, in the Howk et al. (2000) scenario these
emissions arise from failed ejecta returning violently to the star's
surface.  Because the spectrum reveals the presence of the Ly$\alpha$
34\,\AA\ line of \ion{C}{6} (Mewe et al. 2003), for which the
ionization potential is within a factor of 3 or 4 of \ion{O}{6}, it
does seem reasonable to us to assume a connection between the soft
X-ray and UV attributes. This leads to the question: can the warm
k{\it T$_2$} and k{\it T$_3$} components in $\gamma$\,Cas be
identified with the 0.7 and 0.1\,keV components in $\tau$\,Sco? We
believe the answer to this question is ``no."  According to Cohen et
al. and Mewe et al., the $\ge$2.5 and 0.7\,keV component emissions are
likely to arise from the peculiar properties of this star's wind, such
as from returning clumps of a partially stalled wind.  be explained by
$\sim$1000 infalling clumps, which would represent a substantial
fraction of the mass in the star's exosphere at any one time (Mewe et
al. 2003). Yet, in our view it does not seem likely that this scenario
can occur in $\gamma$\,Cas's wind.\footnote{The following discussion
presupposes that the wind of $\gamma$\,Cas is similar to those of
other Be stars for its spectral type, as indeed previous studies have
suggested (e.g., Henrichs et al. 1983, Doazan et al. 1987, Grady et
al. 1987). It is possible that $\tau$\,Sco's is observed from a polar
aspect, which might explain its anomalous strength for a B0
star. Thus, while the emission measures of $\gamma$\,Cas's warm
subcomponents are puzzling on one hand, it is actually the strength of
$\tau$\,Sco's wind that remains the long-standing enigma.  } Consider
that its wind is much weaker than $\tau$\,Sco's, either as measured by
its mass loss rate ($3.1\times10^{-8}M_{\odot}$
vs. $\sim1\times10^{-8}M_{\odot}$; Howk et al. 2000, Waters et
al. 1987) or its terminal velocity (-2400 km\,s$^{-1}$ vs.  -1800
km\,s$^{-1}$; Howk et al. 2000, Smith \& Robinson 1999). In addition,
even though the wind is weaker in $\gamma$\,Cas, the emission measures
of its components 2 and 3 are nearly ten times larger than the
components with these approximate temperatures in $\tau$\,Sco.  There
is no indication of emission or redshifts in the absorptions of the
\ion{N}{5} and \ion{C}{4} line profiles of $\gamma$\,Cas, and an
archival FUSE spectrum taken in the scattered light of $\gamma$\,Cas
shows no emission or absorption components of the \ion{O}{6}
lines. Thus, it seems difficult to reconcile the hypotheses that the
UV line anomalies are related to the X-ray emission and that these
X-ray components are caused by a common mechanism in the two stars. In
sum, it appears that warm components in the $\gamma$\,Cas X-ray
spectrum are unique for this star.  Moreover, we do not have a firm
picture of the warm site geometry.

We turn next to alternative origins for the k{\it T$_4$} component.
In their {\it ROSAT/PSPC} survey of 27 normal B stars in the solar
neighborhood, Cohen, Cassinelli, \& MacFarlane (1997) found that the
soft X-ray emissions of early B stars are consistent with the a wind
shock origin. The observations show temperatures of $\simeq$\,1\,MK
(0.085\,keV) and typical emission measures of $\simeq10^{54}$
cm$^{-3}$. Although this temperature is consistent with our fourth
component, the typical emission measure in field B stars is a factor
of four lower than we found for model M2.  Even in $\tau$\,Sco, with
its dramatic wind signatures in the UV resonance lines, the emission
measure is a factor of two below our value of EM$_{4}$.  An additional
fact to keep in mind is that the wind signatures in the UV resonance
lines indicate that $\gamma$\,Cas's wind characteristics are
consistent with those of other B0.5e main sequence stars (cf. Henrichs
et al. 1983, CSR00). From these considerations, it seems that at most
only part of the k{\it T$_4$} component can arise from processes
associated with a B-star wind.  We hope that a planned {\it XMM/RGS}
spectrum will shed light on the origin of the cooler components,
particularly, k$T_{4}$.

\subsubsection{The Fe anomaly in the hot plasma component }

In our best models we find the Fe abundance to be approximately
0.22$\pm{0.05}$ solar, yet the Fe and other metallic abundances for
the warm components X-ray emissions are close to the solar value.  By
comparison, the Fe abundance determined for B stars from {\it ASCA,
  Chandra,} and {\it XMM} observations are also nearly solar
(0.6--1.0; e.g., Kitamoto et al. 2000). These values are in agreement
with the general slight underabundance of about 30\% for the local
interstellar medium and photospheric abundances for galactic OB
stars. The data on Fe abundances in X-ray emitting plasmas associated
with somewhat abnormal hot stars are still fragmentary. However, if
the example of the Wolf-Rayet star WR\,25, with possible colliding
winds, is typical, then the Fe abundance does not differ from main
sequence values. Moreover, results for the same stars observed with
the HETGS and {\it XMM} or {\it ASCA}, including $\gamma$\,Cas, are
the same within expected errors. The Fe abundance in the hot plasma
component is certainly lower than values found in other upper main
sequence stars, and we will now discuss several possible causes that
can lead to this anomaly.

\noindent {\it Oversimplified modeling:~} 

A number of simplifying assumptions were made in our global fitting 
analysis that are known to affect Fe abundances estimates. These fall 
into two general categories, nonequilibrium physics and a broad DEM with 
an arbitrarily high temperature limit. The general problem in these cases is 
that the temperatures derived from the continuum shape and the Fe ionization 
often  do not agree. Transient ionization can occur in  
in rapidly evolving rarefied plasmas if the dynamical timescale is shorter 
than the recombination timescale. We do not expect transient states to occur 
for the hot plasma associated with $\gamma$\,Cas because this component 
should have a relatively high density (10$^{11}$\,--\,10$^{13}$ cm$^{-3}$), 
according to the analysis of SRC98. 

  The apparent continuum temperature can also be misdetermined by relying 
on an oversimplified high energy model, e.g., by failing to include 
a needed second high energy component and/or (as for some CVs) by 
reflection of hard X-rays from the nearby star. This problem could be 
important for our case if the temperature measured from the Fe line ratio 
were at the low end of a continuous DEM extending to high temperatures.
The continuum at short wavelengths would then be enhanced by contributions 
from higher temperatures and would dilute the Fe line strengths. 
In actuality, the spectrophotometry of the {\it RXTE} and {\it 
BeppoSax} instruments give no hint of a flux excess at high energies. 
The addition of an ultra-high temperature component is also contradicted
by the agreement of the temperatures inferred from the short-wavelength
{\it ratio} of the Fe\,K line strengths.

\noindent {\it  Anisotropic scattering:~} 
For special geometries in optically plasmas anisotropic scattering
can deflect Fe line photons from the observer's line of sight and 
modify its equivalent width. However, the circumstances surrounding
X-ray flares suggests that this is unlikely. 
Flares occur continually and are likely to be distributed roughly evenly 
around the star's azimuthal sectors. The strength of the K fluorescence 
and the fraction of low-to-high absorption columns of the hot component are 
both consistent with the more or less isotropic sprinkling of these sources 
around the star, offer no special geometry that could remove photons from 
the line of sight. Even if a special configuration offered itself at some 
phase, it is doubtful that it would remain for many viewing angles 
associated with our 53\,ks integration.

\noindent {\it Donor star has a low Fe:~}
The determination of Fe abundances in stars with hard X-ray spectra, 
such as CVs, is fraught with perils caused by 
the reflectivity of the hard X-rays from the star's surface and the
variable K-absorption edge of Fe.  
Most available Fe abundance studies from X-ray lines suggest that they 
lie in an range extending from about 0.2 to, more typically, fully solar 
(Ezuka \& Ishida 1999).
Since the abundance we find for $\gamma$\,Cas is at best 
at the lower boundary of this range, it is not reminiscent of chemical
processing in CV material, which may be due to low abundances in the
secondary donor star. Because of the wide separation between the 
component stars of the the $\gamma$\,Cas binary system (Miroshnichenko
et al. 2003), it is unlikely that 
the surface of the primary would be contaminated by material from 
a metal-deficient secondary.

\noindent {\it Donor is the Be star:~} If the Fe-deficient plasma is
provided by the Be star with assumed normal abundances, then the low
abundances might be somehow caused by processes associated with the
high temperature conditions.  The notion that the low Fe abundance
becomes normal when the plasma cools, as in {\it case 2} above,
implies that surplus Fe atoms are somehow created and restored to the
gas as a part of the cooling process, an {\it ad hoc} proposition. The
alternative is that the high temperatures somehow promote the Fe
anomaly. This concept is reminiscent of the famous First Ionization
potential (``FIP") effect in which elements which have a FIP below
10\,eV observed in the slow solar wind, flares, and comparatively
small-scale structures in the solar corona all exhibit abundances
enhanced by a factor of 3--4 relative to those elements with a FIP
above this value (Feldman 1992) in the slow wind, flares, and
comparatively small-scale coronal magnetic structures. The exact
process for this effect is still unknown but probably involves
incomplete Coulombic coupling between ions due to a combination of
gravitational settling and particle-magnetic interactions occurring in
intermediate-density plasma (see Raymond 1999)

   An ``inverse FIP effect," in which the low-FIP elements such as Fe are 
deficient and the high-FIP ones approximately normal, has been by now 
widely observed in the high-resolution EUV and/or X-ray  spectra associated 
with the coronae of the active primaries of RS\,CVn binaries 
(e.g., Brinkman et al. 2001, Drake et al. 2001, Audard et al. 2003). 
For example, the Fe abundance in the corona of the primary
of the prototypical system HR\,1099, as derived from its {\it XMM/RGS} 
spectrum, is only 0.25 solar.
G\"udel et al. (2003) have suggested that the inverse FIP effect is due 
to high-energy electron beams (as evidenced by 
their observed radio gyrosynchronous radio emission) which propagate along
magnetic lines into the star's chromosphere. Low-FIP ions are prevented 
from escaping into the corona by the strong downward-directed electric 
field created by this beam. 
This mechanism squares seemingly perfectly with the picture suggested 
by Robinson \& Smith (2000) in which X-ray flares of $\gamma$\,Cas are 
produced as a consequence of accelerated beams into its upper atmosphere
from field stresses introduced between the rotation rates of the Be star
and its Keplerian disk.  There may be other circumstantial links between 
the inverse FIP effect and a beam-flare mechanism. 
According to G\"udel et al. (2002), the one other star known to harbor 
magnetic co-rotating clouds, the active K dwarf AB\,Dor, shows X-ray
flaring and an inverse FIP effect. These authors have noted that 
the strength of this effect moderates for Fe during flaring events in
this star, and this also seems to be the case for HR\,1099 (Audard et
al. 2001). The moderation of this effect is indeed predicted during 
strong flaring by the G\"udel et al. mechanism.

   As a variant of the ``classical" FIP interpretation, Dr. S. Cranmer 
(private communication) has also pointed out that {\it SOHO} satellite 
observations of the solar corona have indicated a classical FIP-like 
pattern of abundances for gas located in the middle-regions of large 
solar large solar coronal structures called helmet streamers. 
In these central regions matter seems to be the most stable 
and thus least able to be completely mixed (Raymond 1999). 
The abundance pattern introduced is similar to the classical
FIP effect itself, except a metal deficiency bias is introduced
such that the low-FIP ions like Fe are deficient
by a factor of 3 and the high-FIP ions by a factor of 10. 
Interestingly, complete stability
in atmosphere is apparently not needed for this anomaly to develop. In
the solar case, mass circulation caused by chromospheric spicules is
100$\times$ greater than net outflow assumed in chromospheric diffusion
models (Athay 1976). An additional consideration is that total stability
in the solar corona would mean that gravitational diffusion alone would
impose abundance gradients orders of magnitude greater than is observed,
so some type of mixing seems actually to be required, either by kinetic
or magnetic processes (Raymond 1999). In this connection, we note
that on $\gamma$\,Cas the role of surface flares, 
which are confined to small volumes and which occur 
over timescales much shorter than the gravitational settling timescale, 
may be equally irrelevant to any stability requirements as 
chromospheric spicules are in the solar case. 

  Given the commonality of some form of FIP effect in active cool stars, 
it seems that Fe deficiency is becoming almost an expected
hallmark of active magnetic stars, and indeed may even be linked to
particle beams needed for the production of X-ray flares. Thus, the
existence of this effect in $\gamma$\,Cas is additional circumstantial
evidence that the hard X-rays are produced as a byproduct of magnetic field
stresses associated with the Be star and its disk. We should also note
that if the FIP effect is the correct explanation for the Fe deficiency
it strengthens the argument that the warm plasma comes from a reservoir 
of material separate from the hot component {\it (case 1.)}

\subsection{Comparisons with Cataclysmic Variable Spectra }

   The suggestion has been made recently (Kubo et al. 1998, Apparao 
2002) and countered (Robinson \& Smith 2000) that an active Be\,--\,white 
dwarf system is responsible for the hard and luminous X-ray flux of 
$\gamma$\,Cas. Historically, one of the chief reasons for this suggestion 
was been the detection of moderate-strength H- and He-like Fe lines in the
spectrum, indicating emission from a hot, optically thin gas (Murakami et 
al. 1986). In the general binary accretion picture (and considering more 
carefully the likely evolution of the members of this binary), a young
$\gamma$\,Cas binary could evolve to become a Be-neutron star system. 
However, such spectra exhibit nonthermal spectra and weak Fe K lines. 
Moreover, Be\,--\,n.s. systems generally have eccentric orbits and also and 
have a strong likelihood of emitting X-ray pulses, for which extensive 
searches have proved negative. 

  Let us consider instead the accreting white 
dwarf picture, e.g., as suggested by Kubo et al. (1998). Recently published 
{\it Chandra} spectra of several CVs (notably U\,Gem and TX\,Hya) 
in their quiescent states exhibit
both similarities and dissimilarities to the $\gamma$\,Cas spectrum.
The X-ray continuum of these objects is consistent with
a (generally) optically-thin, two-temperature model.
The hotter, and dominant, component typically has a value k{\it T} $\approx$
15--80\,keV (cf. Szkody et al. 2002, Mauche 2002, Mukai et al. 2003). 
The second ``cool" component (0.1--2 keV) is most easily observed
in EUV wavelengths. In magnetic CVs, Roche-overflowing mass is channeled 
to the magnetic poles
of the white dwarf, causing an accelerated column flow to shock as it impacts 
the star's surface (e.g., Warner 1995). In nonmagnetic stars undisrupted
accretion disks extend to the star's surface and turbulent energy in the
shear layer thermalize and radiate X-rays (Patterson \& Raymond 1985). 
In both cases the emergent
flux distribution results from a broad distribution of temperatures
radiating from the hard X-ray to EUV spectral domains. Because these 
components roughly resemble the multiple components we have identified
in our DEM analysis, there is some resemblance between the spectra of 
$\gamma$\,Cas and the CVs, so we examine this further.

   Mukai et al. (2003) has argued for a dichotomous classification of
CVs based on the spectra of the seven of them observed to date by {\it
Chandra}.  One of these subgroups consists of white dwarfs with low or
undetectable magnetic fields and/or low specific mass transfer
rates. The second subgroup, having much stronger magnetic fields,
consists of intermediate-polars with associated high-rate and
channeled mass transfers, and their spectral properties are consistent
only with photoionization-dominated processes.  These authors point
out that the {\it ASCA} and {\it Chandra} spectra of the nonmagnetic
group exhibit emission primarily in lines of Fe\,L-shell ions
(\ion{Fe}{17}--\small{XXIV}) as well as lines of H-like S, Si, Mg, Ne,
O, and N.  The kinematic properties of this group are also similar to
those inferred for $\gamma$\,Cas. For example, their lines indicate a
rest velocity and broadenings of 200\,--\,550\,km\,s$^{-1}$.
Moreover, spectra of well-studied systems like U\,Gem and TX\,Hya also
show density-sensitive diagnostics that indicate $n_{\rm e}$ $\ge$
10$^{14}$ cm$^{-3}$ and/or small dilution factors for EUV/UV
radiation.  However, the nonmagnetic spectra exhibit neither a
substantial K fluorescence strength nor attenuated long-wavelength
continua.  Spectra of the second, magnetic polar group stand in marked
contrast to spectra of either $\gamma$\,Cas or the nonmagnetic
group. For example, their hard continua are best fit with a nonthermal
model.  Their spectra also have weak or undetectable lines of Fe
L-shell ions as well as only weak lines of the H-like ions. In short,
the $\gamma$\,Cas spectrum cannot be reconciled with the Mukai et
al. dichotomy since this spectrum shows hybrid evidence of dominant
collisional processes and yet also exhibits a moderate-strength Fe
fluorescence feature. The $\gamma$\,Cas spectrum is also unique in
requiring two separate Fe abundances and absorption systems.

\section{ Conclusions}

Our previous series of papers discussed in some detail the temporal
variations of the X-rays of $\gamma$\,Cas and their correlations with
various optical and UV spectral indicators and broadband fluxes (see
Smith \& Robinson 2003).  In this study we have found that the
spectral features of this star are eclectic and suggest at least three
different X-ray generation environments. Taken together, the spectral
signatures appear to be just as unique as its temporal and
colorimetric properties. They may be summarized as follows.

First, the HEG/MEG emission is produced by several plasma components,
each of which seems to be thermal and collisionally dominated.  The
primary ``hot" component has a temperature of 10--12 keV and a
distinctly subsolar Fe abundance. This component is responsible for
the emission of both the short-wavelength continuum and the
\ion{Fe}{25} and \small{XXVI} lines.  A subcomponent of the hot
plasma, about three times smaller in volume than the primary
subcomponent, is heavily attenuated at long wavelengths by gas having
a column density of $\approx 10^{23}$ cm$^{-2}$.  Some 10--14\% of the
total emission measure is produced by a ``warm" component. The DEM of
this warm component may be continuous over the range of about 0.1--3
keV (though it seems to require definite peaks), or it may contain the
three discrete components at k{\it T$_2$} = 3, k{\it T$_3$} = 0.37,
and k{\it T$_4$}=0.15.

  Second, although the different Fe abundances for the warm and hot 
emission sites argues that the emissions are not cospatial, it is 
doubtful that they are completely unrelated. For one thing, a
``warm"-component EM of nearly 10$^{55}$ cm$^{-3}$ seems to be atypical 
among B stars. It is more natural to attribute it to the (also unique) hot 
component, for example, by shock heating as the hot sites impact edges of 
the Be disk. Alternatively, the warm and hot components could be somehow
produced by a common external mechanism, such as through heating 
associated with the magnetic reconnections responsible for the hot
flare and canopy subcomponents. However, since these are all 
speculations, we still have no good sense of the geometry or of the
cause(s) of the warm components.

 Third, the X-rays are absorbed by cool material located in at
least two types of structures. The first type affects both the warm and
most of the hot X-ray emission and has a column density of
1\,--\,3 $\times 10^{21}$ atoms cm$^{-2}$. 
This is most probably the stellar wind or outer regions of the decretion 
disk. The second structure has a column density of  
$\sim 10^{23}$ atoms cm$^{-2}$ and is detected primarily by its effects on
a fraction of the hot X-ray component. Some emission from the warm component
is probably also affected by this gas. However, the column mass is so large
that this emission is completely absorbed. This gas is probably also the
source of the Fe and Si K fluorescence features. We have identified this 
absorbing gas as the dense portion of the circumstellar disk and as
such is in close proximity to the sites of hard X-ray production.

  Fourth, we have emphasized the peculiar Fe deficiency in 
the hot plasma component whereas the warm components show 
nearly solar metallic abundances. So far as we are aware, this is a unique 
attribute of an X-ray spectrum of a hot star. 
Thus if the Fe-abundance anomaly turns out to arise 
from a FIP-like effect, it would be the first discovery in a hot star
(for example, it is not present in $\tau$\,Sco; Mewe et al. 2003). 
The effect itself might then become a proxy in
stars for which detection of Zeeman splitting is often impossible.
If these ideas are approximately correct, one 
might expect to see deficiencies of Fe and other low-FIP elements
in other hot astronomical venues.

  We count three attributes of the spectrum that argue for the association
of the X-rays with either the Be star or its disk: (1) the long-wavelength
attenuation of fluxes, (2) the generation of moderate strength Fe and Si
K fluorescence features, and (3) the Fe abundance deficiency, which we
suggest is associated with the FIP effect. In contrast, these attributes 
are hard to reconcile with high-resolution spectra of the well-observed  
CVs to date. Moreover, the spectral characteristics of $\gamma$\,Cas 
do not fit into the context of OB star behavior either. 
In a preliminary survey of line widths among galactic 
O and early B stars, Cohen et al. (2003) found that line widths decrease 
with spectral type from large values 
down to $\sim$250\,km\,s$^{-1}$ at B0.2. Thus, 
viewed in the context of X-ray generation in a magnetic environment, the
larger line widths for $\gamma$\,Cas are unexpected. Nor are the line
profiles consistent with magnetic confinement or high velocity outflow 
either. Alternatively, although it is not a unique interpretation, we 
have suggested that the line broadening is caused by the co-rotation
of circumstellar material or the Keplerian velocities of the inner Be disk.
It is perhaps pertinent to point out 
that in the magnetic dynamo model of RSH02, it is just those conditions
within a stellar radius where stresses might be expected to lead to 
magnetic dissipation and the production of particle beams and X-rays.

  It is our pleasure to thank by Drs. Steve Federman, Ed Jenkins, Koji
Mukai, Tim Kallman, Chris Mauche, and Paula Szkody for very helpful
discussions.  We are also grateful to Dr. Steve Cranmer for his
introducing us to the literature of the FIP effect in solar helmet
streamers.  We also appreciate a number of insightful comments by the
referee. This work has been supported in part by NASA Grants SAO-2019A
and NAG-5-11705.  MFG is supported by NASA through the Chandra
Postdoctoral Fellowship Award program, Number PF01-10014, issued by
the Chandra X-ray Observatory Center, which is operated by the
Smithsonian Astrophysical Observatory under NASA contract NAS8-39073.

\clearpage

\clearpage
 
\begin{table}[ht!]
\begin{center}
\caption{\label{lines}\centerline{Measured line properties } }
\centerline{~}
\begin{tabular}{cccccc}  \hline \hline 

Ion & $\lambda_{th}$ (\AA) & $\lambda_{obs}$ (\AA) &
$\sigma_\lambda$ (m\AA) & Line Flux$^{a}$ & Cont. Flux$^{b}$ \\
 &   &   &   &   & \\
\hline 

Fe XXVI Ly$\alpha$& 1.780 & 1.7816 (14) & 2.3$\pm$2.0 & 10.6$\pm$2.1 & 5.44\\
S XVI Ly$\alpha$& 4.729 & 4.7295 (12) & 9.4$\pm$1.7 & 8.6$\pm$1.2 & 3.96\\
Si XIV Ly$\alpha$& 6.182 & 6.1848 (11) & 8.4$\pm$1.1 & 7.8$\pm$0.5 & 2.96\\
Mg XII Ly$\alpha$& 8.421 & 8.4216 (23) & 14.8$\pm$3.2 & 4.9$\pm$0.7 & 2.10\\
Ne X Ly$\alpha$& 12.134 & 12.1374 (15) & 18.8$\pm$1.9 & 12.6$\pm$1.6 & 1.26\\
O VIII Ly$\alpha$& 18.969 & 18.9756 (67) &32.0$\pm$6.0 & 17.1$\pm$2.6 & 0.49\\
\hline
Fe XXV $r$ &1.8504&$\cdots$&2.9&6.8$\pm$1.9&6.1\\
Fe XXV $i$ &1.8570&$\cdots$&3.0&5.7$\pm$1.8&6.0\\
Fe XXV $f$ &1.8682&$\cdots$&3.0&2.4$\pm$1.5&5.9\\
Fe I K$\alpha_1$ &1.9358&$\cdots$&3.1&3.6$\pm$1.8&6.2\\
Fe I K$\alpha_2$ &1.9398&$\cdots$&3.1&5.0$\pm$1.5&6.2\\
Si XIII $r$ &6.6479&$\cdots$&10.6&2.0$\pm$0.5&2.6\\
Si XIII $i$ &6.6850&$\cdots$&10.6&0.6$\pm$0.5&2.6\\
Si XIII $f$ &6.7395&$\cdots$&10.7&0.9$\pm$0.4&2.6\\
Mg XI $r$ &9.1687&$\cdots$&14.6&0.9$\pm$0.7&1.8\\
Mg XI $i$ &9.2300&$\cdots$&14.7&0.2$\pm$0.4&1.8\\
Mg XI $f$ &9.3136&$\cdots$&14.8&0.0&1.8\\
Ne IX $r$ &13.447&$\cdots$&21.4&4.3$\pm$0.9&0.96\\
Ne IX $i$ &13.551&$\cdots$&21.6&4.5$\pm$1.3&0.96\\
Ne IX $f$ &13.698&$\cdots$&21.8&0.0&0.95\\
Fe XXIV &10.622&$\cdots$&16.9&3.1$\pm$0.6&1.4\\
Fe XXIV &10.663&$\cdots$&17.0&1.9$\pm$0.6&1.4\\
Fe XXIV/XXIII &11.029&$\cdots$&17.6&2.7$\pm$0.6&1.3\\
Fe XXIV &11.266&$\cdots$&17.9&1.6$\pm$0.8&1.3\\
Fe XVII &15.014&$\cdots$&23.9&5.9$\pm$1.7&0.93\\
Fe XVII &15.266&$\cdots$&24.3&1.5$\pm$1.3&0.92\\
O VIII Ly$\beta$&16.006&$\cdots$&25.5&3.1$\pm$1.6&0.74\\
Fe XVII &16.780&$\cdots$&26.7&2.4$\pm$1.2&0.54\\
Fe XVII &17.051&$\cdots$&27.2&3.3$\pm$1.8&0.48\\
Fe XVII &17.096&$\cdots$&27.2&1.8$\pm$1.3&0.47\\
O VII $r$ &21.602&$\cdots$&34.4&7.7$\pm$2.8&0.34\\
O VII $i$ &21.802&$\cdots$&34.7&7.9$\pm$4.0&0.33\\
O VII $f$ &22.097&$\cdots$&35.2&1.3$\pm$1.6&0.31\\
N VII Ly$\alpha$ &24.789&$\cdots$&39.5&6.1$\pm$2.6&0.26\\
\hline
\end{tabular}
\end{center}
\hspace*{-.00in}$^{a}$Line flux in units of 10$^{-5}$ photons cm$^{-2}$ s$^{-1}$  \\
$^{b}$Continuum flux in units of 10$^{-3}$ photons cm$^{-2}$ s$^{-1}$\AA$^{-1}$  \\
\end{table}

\clearpage 
\begin{table}[ht!]
\begin{center}
\caption{\label{fekt} X-ray derived values of k{\it T} and column density }
\centerline{~}
\begin{tabular}{lcrrr} \hline \hline

Author  & Satellite & k{\it T} (keV) & Fe/Fe$_{\odot}$ & 
{\it N}$_{H}$ (cm$^{-2}$) \\
\hline

Murakami et al.  &     Tenma  &   11.7  &      0.3  &  $1.3\times10^{22}$   \\
(1986)
                 &         &   $\pm{0.8}$  &  $\pm{0.1}$  &          \\
 Parmar et al.    &    EXOSAT  &    8.2  &     0 .26  & $1.5\times10^{21}$  \\
(1993)           &           & $\pm{0.09}$  &   $\pm{0.5}$      &       \\

Kubo et. al.     &    ASCA   &   10.7   &    0.35   &  $1.5\times10^{21}$ \\
(1998)           &           &  $\pm{0.6}$  &  $\pm{0.08}$  & $\pm{0.1}$  \\

SRC98            &  RXTE     & 10.5,11.8    &   0.35   &         $\cdots$   \\
                 &           &  $\pm{0.4}$  &   $\pm{0.1}$   &    \\

RS00             &  RXTE     & 10.8-11.4  &   0.34   &    $2\times10^{22}$  \\
                 &           &   $\pm{0.2}$   &  $\pm{0.1}$  &       \\

Owens et al.     &    BpSax  &  12.3    &   0.42  &  $1.55\times^10{21}$  \\
(2001)           &           &  $\pm{0.6}$  &  $\pm{0.05}$  &  $\pm{0.09}$ \\

This paper & Chandra & --  & 0.22\,-\,.26 &  $5\times^10{21}$ \\

\hline
\end{tabular}
\end{center}
\end{table}


\begin{table}[ht!]
\begin{center}
\caption{\label{params} Fit Parameters }
\centerline{~}
\begin{tabular}{rcccc} \hline \hline
 Parameters & M$_1$        &         M$_2$   & M$_3$ & Errors \\
\hline
k{\it $T_1$} (keV)& 12.3 & {\bf 12.3} & 12.3 & -- \\
NORM$_1$\tablenotemark{a} & 0.112 & {\bf 0.134} & 0.135 & $\pm{.022}$ \\
$EM_1$ ($10^{54}$\,cm$^3$)\tablenotemark{b} & 47 & {\bf 57} & 57 & $\pm{10}$ \\
k{\it $T_2$} (keV) & 1.86 & {\bf 3.08} & 2.50 & $\pm{1.2}$ \\
NORM$_2$ & 0.0071 &  {\bf 0.011}  & 0.012 & $\pm{.004}$ \\
$EM_2$ ($10^{54}$\,cm$^3$)& 3.0 & {\bf 4.6} & 5.1 & $\pm{1.6}$ \\
k{\it $T_3$} (keV) & 0.370 & {\bf 0.375} &  0.393 & $\pm{.005}$ \\
NORM$_3$ & 0.0032  & {\bf 0.0023} & 0.0026 & $\pm{.0009}$ \\
$EM_3$ ($10^{54}$\,cm$^3$)& 1.4 & {\bf 1.0} & 1.1 & $\pm{.4}$ \\

k{\it $T_4$} (keV)& 0.135 & {\bf 0.146} & 0.148 & $\pm{.011}$ \\
NORM$_4$ & 0.014  & {\bf .0088}  &  0.0092 & $\pm{.005}$ \\
$EM_4$ ($10^{54}$\,cm$^3$)& 5.9 & {\bf 3.7} & 3.9 & $\pm{2.2}$ \\
ABUND$_{\mbox{Fe}}$\tablenotemark{c} & 0.34 & {\bf 0.22} & 0.26 & $\pm{.05}$ \\
ABUND & 1.1  & {\bf 0.81} & 0.77 & $\pm{0.3}$ \\
{\it N}$_H$ ($10^{21}$\,cm$^{-2}$)\tablenotemark{d} & 3.7 & {\bf 2.7} & 2.7&$\pm{1.0}$ \\
\hline
\end{tabular}
\end{center}

\tablenotetext{a}{For M2 and M3, NORM1 is the total normalization of the 
two hot subcomponents, each having a different absorption.}
\tablenotetext{b}{A Hipparcos distance of 188 pc is assumed.}
\tablenotetext{c}{For M1 and M2, ABUND$_{\mbox{Fe}}$
is the iron abundance of the hot
components alone. For M3, this is the iron abundances of both hot and warm
components, which are tied together.}
\tablenotetext{d}{For M1 this is the column density for all components,
for M2 and M3, the column density for the k$T_{\rm 2}$, k$T_{\rm 3}$, k$T_{\rm 4}$
component and 75\% of the hot subcomponent k$T_{\rm 1}$.}
\end{table}

\end{document}